\documentclass[aps,prl,twocolumn,superscriptaddress]{revtex4-1}
\usepackage[english]{babel}
\usepackage[dvips]{graphicx}
\usepackage{dcolumn}
\usepackage{bm}
\usepackage{graphicx,color}
\usepackage{braket}
\usepackage{amsmath}
\usepackage{etaremune}
\usepackage{layouts}
\usepackage{tabularx}
\usepackage{siunitx}
\usepackage{multirow}
\usepackage[acronym,shortcuts]{glossaries}

\newacronym[longplural={double quantum dots}]{DQD}{DQD}{double quantum dot}

\newcommand{\Jint}{\ensuremath{J_{\textrm{int}}}}

\begin{document}

\title{Phonon spectral density in a GaAs/AlGaAs double quantum dot}

\author{A. Hofmann}
\email{andrea.hofmann@alumni.ethz.ch}
\affiliation{Solid State Physics Laboratory, ETH Zurich, CH-8093 Zurich, Switzerland}
\affiliation{Institute of Science and Technology Austria, Am Campus 1, 3400 Klosterneuburg, Austria}
\author{C.~Karlewski}
\affiliation{Institut f\"ur Theoretische Festk\"orperphysik, Karlsruhe Institute of Technology, DE-76131 Karlsruhe, Germany}
\affiliation{Institute of Nanotechnology, Karlsruhe Institute of Technology, DE-76344 Eggenstein-Leopoldshafen, Germany}
\author{A.~Heimes}
\affiliation{Institut f\"ur Theoretische Festk\"orperphysik, Karlsruhe Institute of Technology, DE-76131 Karlsruhe, Germany}
\author{C.~Reichl}
\affiliation{Solid State Physics Laboratory, ETH Zurich, CH-8093 Zurich, Switzerland}
\author{W.~Wegscheider}
\affiliation{Solid State Physics Laboratory, ETH Zurich, CH-8093 Zurich, Switzerland}
\author{G.~Sch\"on}
\affiliation{Institut f\"ur Theoretische Festk\"orperphysik, Karlsruhe Institute of Technology, DE-76131 Karlsruhe, Germany}
\affiliation{Institute of Nanotechnology, Karlsruhe Institute of Technology, DE-76344 Eggenstein-Leopoldshafen, Germany}
\author{K.~Ensslin}
\affiliation{Solid State Physics Laboratory, ETH Zurich, CH-8093 Zurich, Switzerland}
\author{T.~Ihn}
\affiliation{Solid State Physics Laboratory, ETH Zurich, CH-8093 Zurich, Switzerland}
\author{V.~F.~Maisi}
\affiliation{Solid State Physics Laboratory, ETH Zurich, CH-8093 Zurich, Switzerland}
\affiliation{NanoLund and the Department of Physics, Lund University, Box 118, 22100 Lund, Sweden}
\email{ville.maisi@ftf.lth.se}


\date{\today}
\begin{abstract}
We study phonon emission in a GaAs/AlGaAs double quantum dot by monitoring the tunneling of a single electron between the two dots. We prepare the system such that a known amount of energy is emitted in the transition process. The energy is converted into lattice vibrations and the resulting tunneling rate depends strongly on the phonon scattering and its effective phonon spectral density. We are able to fit the measured transition rates and see imprints of interference of phonons with themselves causing oscillations in the transition rates.
\end{abstract}

\maketitle

\section{Introduction}
Quantum dots in semiconductors constitute a basic building block for a vast variety of experiments due to their discrete energy level structure and high level of tunabiltity. Charge and spin states in single and double quantum dots are candidates for quantum bits~\cite{loss_quantum_1998,imamoglu_quantum_1999}. \Glspl{DQD} are being tested as coherent single photon emitters~\cite{liu_photon_2014,gullans_phonon-assisted_2015,stockklauser_microwave_2015} as well as for their applicability as photon to electron spin converters~\cite{oiwa_conversion_2017}. All of these experiments suffer from relaxation and dephasing of quantum states due to their interaction with the environment. Decoherence of spin states occurs, for example, due to the randomly fluctuating magnetic field produced by the nuclear spins in the host material or due to the spin--orbit interaction coupling electronic noise to the spin degree of freedom~\cite{hanson_spins_2007}. It is possible to reduce spin decoherence due to the nuclear hyperfine effect by polarizing the nuclear spins and narrowing their distribution~\cite{burkard_coupled_1999,coish_hyperfine_2004,petta_dynamic_2008,vink_locking_2009,bluhm_enhancing_2010}. The spin--orbit coupling can be minimized by a suitable alignment of the quantum dots with respect to the crystal and the external fields~\cite{golovach_spin_2008,stepanenko_singlet-triplet_2012,nichol_quenching_2015,hofmann_anisotropy_2017}. Charge noise, on the other hand, is much more difficult to minimize and plays a role for almost any solid-state qubit. Decoherence due to electron-phonon coupling can affect both spin and charge decoherence and persists at zero temperature via phonon emission. Therefore, a detailed study of this effect is in order.
\begin{figure}[t]
	\centering
	\includegraphics{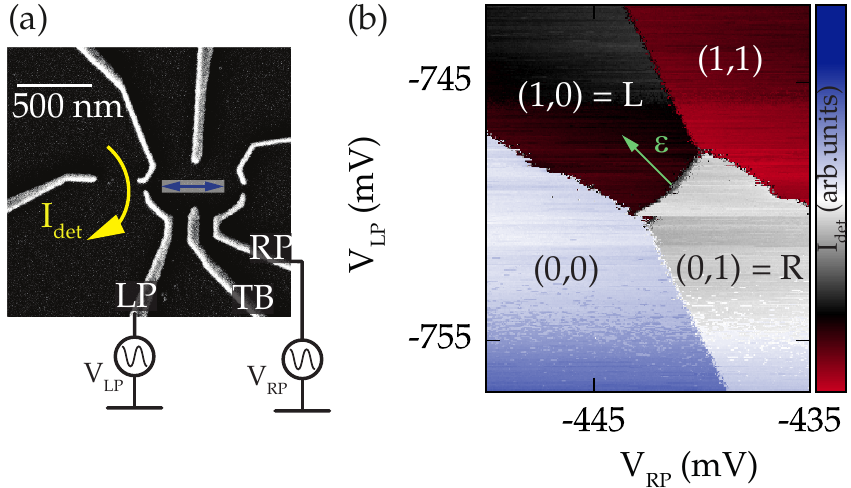}
	\caption{(a) A scanning electron micrograph of our \gls{DQD}. Bright grey fingers display metallic top gates. A single electron tunnels between the two dots as illustrated with the blue arrow. A quantum point contact is formed on the left side of the \gls{DQD} and by measuring the current $I_\mathrm{det}$ (yellow) we determine the electron occupancy of each dot. We control the occupation and energy of the electron states of the two dots with gate voltages $V_{LP}$ and $V_{RP}$ applied to the plunger gates RP and LP, respectively. Gate TB tunes the tunnel coupling between the dots. The \gls{DQD} is isolated from the lead reservoirs. (b), The detector current as a function of $V_{LP}$ and $V_{RP}$ around the transition $(1,0) \leftrightarrow (0,1)$. A linear contribution to $I_\mathrm{det}$ arising from the cross-capacitances between the plunger gates and the dots is subtracted from the signal. We used a bias voltage of $200\ \mathrm{\mu V}$ applied to the QPC and the current level was around $I_\mathrm{det} \sim \SI{6}{\nano\ampere}$.}
	\label{fig:Exp}
\end{figure}


The coupling strength of an electron to phonons depends on the type of interaction as well as on the energy and momentum of either particle~\cite{brandes_dicke_2000}. Describing the energy-dependence of the electron-phonon coupling as a simple form of a spectral density~\cite{brandes_spontaneous_1999} is challenging in semiconductor quantum dots, due to their discrete energy level structure. In pioneering experiments, Fujisawa~\textit{et al.}~\cite{fujisawa_spontaneous_1998} measured inelastic contributions to current running through a GaAs \gls{DQD}  which they attributed to acoustic phonons. They found the current to depend on the detuning $\varepsilon$ as $1/\varepsilon^s$ with $s$ between $1$ and $2$. Further experiments \cite{gasser_statistical_2009,weber_probing_2010,roulleau_coherent_2011,granger_quantum_2012} and theoretical studies \cite{brandes_spontaneous_1999,brandes_dicke_2000,golovach_phonon-induced_2004,brandes_coherent_2005,golovach_spin_2008} confirmed the importance of electron-phonon interaction in semiconductor quantum dots. Nevertheless, a detailed study of the interaction spectral density $\Jint (\varepsilon)$ in a simple, isolated system has been missing.

Our experiment is designed to accurately measure the energy-dependence of phonon-emission caused by electron tunneling events between two quantum dots forming a \gls{DQD}. The energy of the emitted phonon is well-defined and precisely controlled by the energy difference between the quantum states of the individual dots. Using only a single electron in the device makes the system as simple as possible and averts potential complication arising from spin physics~\cite{ono_current_2002} or different degeneracies of the quantum dot states~\cite{hofmann_measuring_2016,tarucha_shell_1996,hofmann_measuring_2016,kurzmann_optical_2016}. In contrast to earlier experiments mentioned above, in our system, the discrete states of interest are well separated from the electronic reservoirs. Also, the tunneling events of the single electron from one dot to another are well separated in time and each event is detected individually. The tunable tunneling coupling between the two dots determines the tunneling rate. This experimental setup will allow us to accurately determine $\Jint (\varepsilon)$.

The outline of this paper is as follows: we first introduce the details of our measurement setup, followed by the presentation of our main experimental results. We review the theory derived in Ref.~\citealp{brandes_spontaneous_1999} and employ the interaction spectral density $\Jint$ of electron-phonon coupling according to Ref.~\citealp{gasser_statistical_2009}. This provides a basis to compare the relative strength of different coupling mechanisms and to discuss the peculiarities and importance of the electron-phonon interaction, such as phonon-interference terms and induced decoherence.


\section{Experiments}
\label{cExp}
We perform our experiments in a \gls{DQD} formed in a two dimensional electron gas at a GaAs/AlGaAs interface $\SI{90}{\nano\meter}$ below the surface. The device is shown in Fig.~\ref{fig:Exp} (a). Negative voltages applied to the metallic top gates (bright fingers) deplete the underlying electron gas. In this way, we define a \gls{DQD} containing a single electron~\cite{tarucha_shell_1996,ciorga_addition_2000} as well as a quantum point contact (QPC). The current $I_\textrm{det}$ through the QPC detects tunneling events between the \gls{DQD} and the source and drain reservoirs~\cite{vandersypen_real-time_2004,schleser_time-resolved_2004} as well as in-between the two dots~\cite{fujisawa_bidirectional_2006}. As a function of voltage applied to the left (LP) and right (RP) plunger gates, the current $I_\textrm{det}$ shown in Fig.~\ref{fig:Exp}(b) is constant in regions of stable charge and exhibits steps when the ground charge state of the \gls{DQD} changes. Pairs of numbers \textbf{$(N_L,N_R)$} denote the ground state occupation of the \gls{DQD} with $N_{L/R}$ the number of electrons in the left/right dot, respectively. We limit our experiment to the green line indicated in the graph. Along that line, only the $(1,0)$ and $(0,1)$ states are populated and tunneling is merely allowed in-between the two dots. Tunneling to source or drain reservoirs is energetically suppressed by a high charging energy of the order $\SI{1}{\milli\electronvolt}$~\cite{ciorga_addition_2000,hofmann_measuring_2016}. 

\begin{figure}[t]
	\centering
	\includegraphics[width=0.4\textwidth]{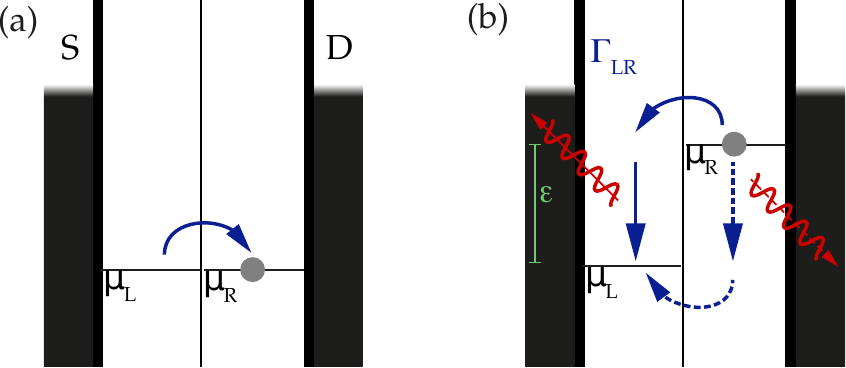}
	\caption{Panel (a) illustrates the preparation of the \gls{DQD} system, where the chemical potentials $\mu_L$ and $\mu_R$ are in resonance. The detection of the electron in the right dot triggers the detuning of the energy levels by an energy $\varepsilon$. The detuning is shown in green in panel (b). In this situation, an electron may tunnel to the left dot, thereby emitting a phonon at $\hbar \omega_{q}$.
	\label{fig:Sketch}}
\end{figure}

We measure energy-dependent tunneling rates between the two dots by employing a feedback loop as in Ref.~\citealp{hofmann_measuring_2016}. The $(0,1)$ and $(1,0)$ states are first tuned into resonance as indicated in Fig.~\ref{fig:Sketch}~(a). Once the electron resides in the right dot, we lift the energy level $\mu_\textrm{R}$ of the right dot by decreasing the voltage $V_{RP}$, and we lower $\mu_\textrm{L}$ by increasing $V_{LP}$ by the same amount (cf. green arrow in Fig.~\ref{fig:Exp}~(b)). Knowing the conversion factor between gate voltages and quantum dot energies from finite bias measurements~\cite{ihn_semiconductor_2010}, the voltage difference defines the energy difference $\varepsilon=\mu_\textrm{R}-\mu_\textrm{L}$ between the two dots. An electron tunneling from the right to the left dot dispenses this energy $\varepsilon>0$ to the environment. Having detected the tunneling event, we bring the $(0,1)$ and $(1,0)$ back in resonance and start the cycle from the beginning. Every repetition provides an instance of the electron waiting time in the right dot, which finally allows us to determine the tunneling rate $\Gamma_{LR}(\varepsilon)$ as the inverse of the mean waiting time~\cite{vandersypen_real-time_2004,schleser_time-resolved_2004,hofmann_measuring_2016}. Similarly, we measure $\Gamma_{RL}(\varepsilon)$ at $\varepsilon<0$, when energy is released as an electron tunnels from the energetically higher left to the lower right dot state.

\begin{figure}[tb]
	\includegraphics[width=0.48\textwidth]{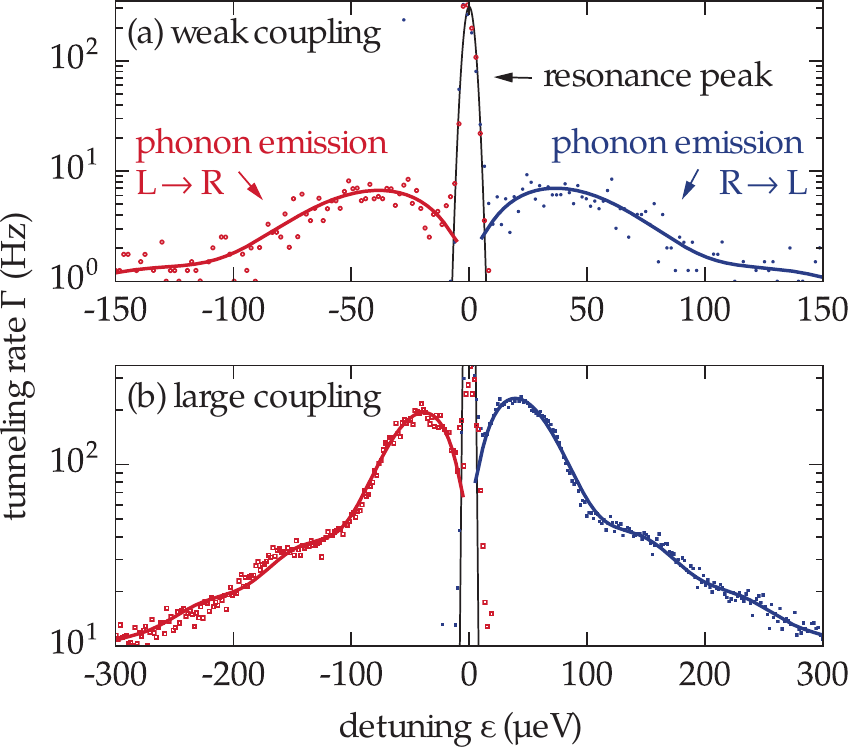}
	\caption{Energy-dependent tunneling rates $\Gamma_{\mathrm{LR}}(\varepsilon)$ ($R\rightarrow L$, blue filled circles) and $\Gamma_{\mathrm{RL}}(\varepsilon)$ ($L\rightarrow R$, red open squares). Around zero detuning, a Gaussian lineshape (black solid line) fits the data. The solid blue and red lines denote fits to Eq.~\eqref{eq:Gamma} of the data measured at larger detuning, $|\varepsilon|>\SI{10}{\micro\electronvolt}$. The electronic temperature of our system, $T=\SI{50}{\milli\kelvin}$ (measured from finite-bias Coulomb resonances), is significantly lower than the detuning $\varepsilon = \SI{40}{\micro\electronvolt}$, at the phonon emission peak.}
	\label{fig:Fit}
\end{figure}

The measured tunneling rates are shown in Fig.~\ref{fig:Fit} in blue for $\Gamma_{LR}(\varepsilon)$ and in red colour for $\Gamma_{RL}(\varepsilon)$, respectively. At low detuning we observe a resonance peak which is much wider than expected for the small tunneling coupling at hand and which we therefore attribute to random fluctuations of the electrostatic environment~\cite{maisi_spin-orbit_2016} and to which we therefore fit a Gaussian lineshape (black solid line). Its full width at half maximum amounts to \SI{5}{\micro\electronvolt}. The value of gate voltage where the peak is maximum defines the zero energy reference, $\varepsilon=0$.

In addition to the resonance peak, we observe finite tunneling rates at $\varepsilon > \SI{10}{\micro\electronvolt}$, for tunneling from the right dot to the left and at $\varepsilon < \SI{10}{\micro\electronvolt}$, for tunneling from the left to the right quantum dot. In both cases, energy is released to the environment and converted to phonons similarly as reported in~\cite{fujisawa_spontaneous_1998,brandes_spontaneous_1999}. To increase the phonon emission rate, we enlarge the tunnel-coupling (ie. increase the voltage applied to gate TB). As shown in Fig.~\ref{fig:Exp}(b), the better statistics reveals oscillations in the tunneling rate at large emission energies $|\varepsilon|>\SI{100}{\micro\electronvolt}$. Tunneling events from the energetically lower to the higher quantum dot states (phonon absorption) are suppressed, as expected for the bath temperature of $\SI{50}{\milli\kelvin}$.


\section{Theoretical Model}
\label{cTheo}
Following Ref.~\citealp{brandes_spontaneous_1999,brandes_dicke_2000}, we now recapitulate the theoretical model relating the observed tunneling rates of Fig.~\ref{fig:Fit} to the spectral density $\Jint (\varepsilon)$ of electron-phonon coupling. The Hamiltonian $H = H_{el} + H_{ph} + V_{el-ph}$ describing our system consists of the energy term for the electron in the \gls{DQD},
\begin{align}
 H_{el}=\frac{\varepsilon}{2}(c_L^\dagger c_L-c_R^\dagger c_R)+\frac{t}{2}(c_L^\dagger c_R+c_R^\dagger c_L), 
\end{align}
the term describing the phonon energy,
\begin{align}
 H_{ph}=\sum_q\hbar \omega_q a_q^\dagger a_q,
\end{align}
and the coupling between the electrons and the phonons,
\begin{align}
\label{eEPh}
 V_{el-ph}=&\sum_{r=L,R}\sum_q \int dx\notag\\&\times V(x)\phi_r^*(x)\phi_r(x)e^{iqx}c_r^\dagger c_r(a_q+a_{-q}^\dagger).
\end{align}
In the first term, $\varepsilon$ describes the detuning, $t$ the tunnel coupling between the two dots, and the operators $c_i$ and $c_i^\dagger$ respectively annihilate and create an electron in the left ($i=L$) or right ($i=R$) dot. Similarly, the operators $a_q$ and $a_q^\dagger$ in the second term correspondingly annihilate and create a phonon with momentum $q$ and frequency $\omega_q$ (energy $\hbar \omega_q$). In the third term, the generic potential $V(x)$ includes deformation and piezoelectric electron-phonon coupling. The spatial $x$-axis traverses both, the left and right quantum dot at positions $x_{L,R}$, respectively, and its origin lies at half the distance between them. Finally, $\phi_i(x)$ are the electronic wave functions of dots $i=L,R$.

We introduce the Fourier transform $\lambda_q^r=\int dxV(x)\phi_r^*(x)\phi_r(x)e^{iqx}$ to write
\begin{align}
 V_{el-ph}=\sum_{r=L,R }\sum_q \lambda_q^rc_r^\dagger c_r(a_q+a_{-q}^\dagger),
\end{align}
Assuming that the electronic wave functions in the left and right dot are equal up to a shift in real space and a global phase, i.e. $\phi_{L/R}=\phi_e(x-x_{L/R})$, yields the canonical transformation
\begin{align}
 \tilde{H}= &\ e^{iS}He^{-iS}\approx H+i[S,H]\notag\\= &\ H_0+V_{el-ph}+i[S,H_0]+i[S,V_{el-ph}], 
\end{align}
with
\begin{align}
 S=\sum_{rq}\frac{\lambda_q^r}{\hbar \omega_q}c_r^\dagger c_r(a_q-a_{-q}^\dagger).
\end{align}
Neglecting terms of higher order in $\lambda_q^r/\hbar \omega_q$ as justified for weak electron-phonon coupling leads to
\begin{align}
\label{eq:HamiltonianSimple}
 \tilde{H}=&\left(\frac{\varepsilon}{2}-\sum_q\frac{|\lambda_q^L|^2}{\hbar\omega_q}\right)c_L^\dagger c_L+\left(-\frac{\varepsilon}{2}-\sum_q\frac{|\lambda_q^R|^2}{\hbar\omega_q}\right)c_R^\dagger c_R\notag\\
 &+\sum_q \hbar\omega_qa_q^\dagger a_q +\left\{\frac{t}{2}c_L^\dagger c_R\left[1-\frac{\lambda_q^L-\lambda_q^R}{\hbar\omega_q}(a_q-a_{-q}^\dagger)\right]\right.\notag\\
 &\left.\vphantom{\frac{\lambda_q^L-\lambda_q^R}{\omega_q}}+h.c.\right\}.
\end{align}
Now, Fermi's golden rule applied to the last term provides the phonon emission (e) and absorption (a) rates
\begin{align}
\label{eq:gFermi}
 \Gamma^a(\varepsilon>0)=&\ \frac{t^2}{8\pi\hbar}\sum_q\frac{\left|\lambda_q^L-\lambda_q^R\right|^2}{(\hbar\omega_q)^2}n_q\delta(\varepsilon-\hbar\omega_q),\\
 \Gamma^e(\varepsilon<0)=&\ \frac{t^2}{8\pi\hbar}\sum_q\frac{ \left|\lambda_q^L-\lambda_q^R\right|^2}{(\hbar\omega_q)^2}[1+n_q ]\delta(\varepsilon-\hbar\omega_q),
\end{align}
respectively. The previous assumptions imply that $\lambda_q^L=\lambda_q=\lambda_q^Re^{-iqd}$ with $d$ the distance between the dots, which allows for the definition of the interaction spectral density
\begin{align}
 \Jint (\epsilon)&=\sum_q|\lambda_q|^2|1-e^{iqd}|^2\delta(\epsilon-\hbar\omega_q).
\end{align}
The spectal density of piezoelectric interactions as well as interactions with bulk deformation phonons have been discussed in detail in Refs.~\cite{brandes_dicke_2000,gasser_statistical_2009}. Here we employ the explicit energy dependence~\cite{gasser_statistical_2009}
 \begin{align}
 \label{eq:SpectralDensity}
 & \Jint (\epsilon) = \gamma \Delta^s e^{-(\Delta r_0/\hbar c)^2} \left[h\left(0\right)-h\left(\frac{\Delta d}{\hbar c}\right)\right],\\
 &\Delta=\sqrt{\epsilon^2+4t^2} \rightarrow \epsilon \textrm{, for } |\epsilon|\gg t.
\end{align}
The exponential dependence on the radius of a single dot, $r_0$, signifies the suppression of coupling to phonons with wave-vectors larger than the inverse radius. Table~\ref{tab:ParametersThomas} provides the coupling parameters $\gamma$ expected in GaAs, the exponent $s$ and the functions $h(\eta)$ for coupling to piezoelectric longitudinal (pe,L), piezoelectric transversal (pe,T) and deformation potential phonon modes. The speed of sound is $c=\SI{5e3}{\meter/\second}$ in GaAs~\cite{madelung_gallium_2002}.
\begin{table*}[t]
	\caption{Electron-phonon interactions.}
	\begin{tabularx}{\linewidth}{X X X c}
		\toprule
		Phonon & $\gamma (\si{\electronvolt^{-s}\second^{-1}})$ & s & $h(\eta)$ \\
		\colrule & & & \vspace{-2ex} \\
		pe,L & $\SI{5e13}{}$ & $1$ & $-72\eta^{-7}\left[9\eta(\eta^2-10)\cos(\eta)+(\eta^4-39\eta^2+90)\sin(\eta)\right]$ \\
		\vspace{1ex} & & & \\
		pe,T & $\SI{5e13}{}$ & $1$ & $-16\eta^{-7}\left[\eta(\eta^4-51\eta^2+405)\cos(\eta)-3(3\eta^4-62\eta^2+135)\sin(\eta)\right]$ \\
		\vspace{1ex} & & & \\
		dp    & $\SI{7.3e20}{}$ & $3$ & $2\eta^{-1}\sin(\eta)$ \\
		\botrule
	\end{tabularx}
	\label{tab:ParametersThomas}
\end{table*}

The phonon emission rate caused by electron tunneling between the two quantum dot levels at large absolute values of the detuning compared to the tunnel coupling and the temperature, $|\epsilon|\gg t,T$, reads
\begin{align}
	\label{eq:Gamma}
	\Gamma^e(\epsilon<0) = \frac{t^2}{\epsilon^2}
	\Jint (\epsilon).
\end{align}
In these conditions, which are met in our experiments, the electron tunneling rate gives a direct measure of the interaction spectral density.

\section{Discussion}
\subsection{Quantification of interaction strengths}
In order to estimate the contributions of different coupling modes and mechanisms we fit our data
\begin{align}
\label{eq:FitThomas}
\Gamma^{e}_\textrm{tot}=&a_\textrm{PE,L}\Gamma^{e}_\textrm{pe,L}+a_\textrm{PE,T}\Gamma^{e}_\textrm{pe,T}+a_\textrm{DP}\Gamma^{e}_\textrm{dp}.
\end{align}
where the observed electron-phonon coupling is described by a contribution of three terms: two piezoelectric terms (parameter $s=1$) and a deformation potential term ($s=3$). Each individual term is described by Eq.~\ref{eq:Gamma} with the corresponding parameters given in Tab.~\ref{tab:ParametersThomas}.

For simplicity, we assume equal speed of sound for longitudinal and transversal modes, hence equal coupling to piezoelectric longitudinal and transversal modes, $a_\textrm{PE} \overset{\textrm{def}}{=} a_\textrm{PE,T}=a_\textrm{PE,L}$. We therefore use as fitting parameters the two dimensionless amplitudes $a_\textrm{PE}$ and $a_\textrm{DP}$ as well as the distance $d$, while assuming $kT\ll \varepsilon$ and $r_0=\SI{20}{\nano\meter}$. In order to focus on the phonon emission, we omitted data points at $|\varepsilon|<\SI{10}{\micro\electronvolt}$ around the resonance. In this regime, the condition $|\epsilon|\gg t$ is fulfilled, as we will argue below that $t=\SI{15}{\nano\electronvolt}$ for the case of "weak coupling" and $t=\SI{80}{\nano\electronvolt}$ for "large coupling" between the two dots forming the \gls{DQD}.

The fit produces values for the inter-dot distance $d$ ranging from \SI{251}{\nano\meter} to \SI{271}{\nm} with error bars smaller than \SI{5}{\percent} for each of the four data sets. 
The effective strength of piezoelectric coupling is $(\gamma_\textrm{pe,L}+\gamma_\textrm{pe,T}) a_\textrm{PE} = \SI{1.6e12}{\electronvolt^{-1}\second^{-1}}$, where the error to $a_\textrm{PE}$ is \SI{1}{\percent}. The fit accuracy for $a_\textrm{DF}$ is $\SI{30}{\percent}$, and hence we calculate an upper bound for the deformation potential coupling strength, $\gamma_\textrm{dp} a_\textrm{DF} = \SI{7.3e18}{\electronvolt^{-3}\second^{-1}}$. For the typical energy scale of $\epsilon=\SI{100}{\micro\electronvolt}$ we access in our measurements, the piezoelectric coupling is $\SI{1.6e8}{\second^{-1}}$ and therefore almost two orders of magnitude larger than the deformation potential coupling amounting to $\SI{7.3e6}{\second^{-1}}$.

The model of Eq.~\eqref{eq:FitThomas} with the described parameters result in fits as shown in Fig.~\ref{fig:Fit} with solid lines. While it reproduces our data, it is sensitive to the distance $r_0$. The fitted relative contribution $a_\textrm{PE}$ increases for lower values of $r_0$, while the quality of the fit measured by the $\chi$-square tends to be poorer. Requiring purely piezoelectric coupling (fixing $a_\textrm{DP}=0$) gives good agreement at low energies $|\varepsilon|<\SI{120}{\micro\electronvolt}$ but underestimates the tunneling rates at high energies, fulfilling the expectation that deformation potential is more dominant at higher energies.

%
%


\subsection{Interference}
Interference terms in the electron-phonon coupling have been reported already in earlier works~\cite{fujisawa_spontaneous_1998,brandes_spontaneous_1999,weber_probing_2010,roulleau_coherent_2011,granger_quantum_2012}. The difference here is the complete decoupling of the double quantum dot system from the electronic reservoirs. The oscillations in the tunneling rate in this regime substantiate the assumption that it is indeed the phonon emission which causes the inelastic tunneling process~\cite{brandes_dicke_2000}. The interference can be understood in the following way. Two possibilities exist for the electron to tunnel from the higher to the lower energy state via emitting a phonon with energy $\varepsilon$ [cf.~Fig.~\ref{fig:Sketch}]: either, it first tunnels and then relaxes (i.e. emits a phonon), or it first relaxes and then tunnels. Positive interference is possible if the emitted phonon has a wavelength $q$ which is commensurate with the distance between the two quantum dots. Mathematically, this is captured in the geometry factor $h(0)-h(\Delta d/ \hbar c)$ in Eq.~\eqref{eq:SpectralDensity}. The fitted value $d=\SI{260}{\nano\meter}$ for the inter-dot distance agrees with estimations of $d$ from the scanning electron micrograph in Fig.~\ref{fig:Exp}(a).

\subsection{Decoherence}
Phonons are expected to limit the fidelity of charge qubits in \gls{DQD}s~\cite{petta_manipulation_2004,toida_vacuum_2013,stockklauser_strong_2017,mi_coherent_2017} as well as the coherence in photon emission experiments~\cite{liu_photon_2014,gullans_phonon-assisted_2015,stockklauser_microwave_2015}. Based on our results obtained at weak inter-dot tunnel coupling, we estimate the phonon mediated tunneling rate $\Gamma^{e}$ relevant for charge qubits, where typically $t\approx\SI{50}{\micro\electronvolt}$~\cite{petta_manipulation_2004,toida_vacuum_2013,liu_photon_2014,gullans_phonon-assisted_2015,stockklauser_microwave_2015}. For this, we first determine the tunneling coupling of our \gls{DQD} and then use the relation $\Gamma^{e}\propto t^2$ [cf.~Eqs.~\eqref{eq:gFermi}, and text below] to extract the phonon absorption rate at different couplings.

We use the resonance around zero detuning to determine $t$. Without requiring knowledge about details of the coupling mechanism in this regime, we use Fermi's golden rule via Ref.~\citealp{ingold_charge_1992} to write $\Gamma(\varepsilon)=\frac{2 \pi}{\hbar} t^2 P(\varepsilon)$, with $P(\varepsilon)$ the probability to exchange energy $\varepsilon$ with the relevant environment~\cite{ingold_charge_1992}. Integrating both sides of the equation in energy provides
\begin{equation}
\label{eq:tPE}
t^2 = \frac{\hbar}{2 \pi} \int \Gamma(E) dE.
\end{equation}
We integrate the experimental data in Fig.~\ref{fig:Fit}(a) around resonance and obtain $t=\SI{15}{\nano\electronvolt}=h\cdot\SI{3.6}{\mega\hertz}$ for the "low coupling" setting [panel (a)]. From that, we can estimate the tunnel coupling in the "high coupling" regime, where integrating over the measured data is not possible due to the limited bandwidth. Knowing that the phonon emission rate is a factor of \num{30} larger in panel~(b), we conclude that the tunnel coupling is larger by a factor of $\sqrt{30}$, hence $t=\SI{80}{\nano\electronvolt}$ for our experiment at "high coupling".

We now use the phonon emission rate of $\SI{200}{\hertz}$ measured in panel (b) to calculate the emission rate for typical qubit energies. Explicitly, we find $\Gamma^{e}=(\frac{\SI{50}{\micro\electronvolt}}{\SI{80}{\nano\electronvolt}})^2\cdot\SI{200}{\hertz}=\SI{80}{\mega\hertz}$ as the phonon-mediated tunneling rate between two strongly coupled quantum dots. This value matches the observed dephasing rates of state-of-the-art GaAs charge qubits~\cite{toida_vacuum_2013,stockklauser_microwave_2015,stockklauser_strong_2017}, suggesting that coupling to phonons plays an important role. Possibilities to reduce electron-phonon coupling could include using lower-dimensional systems, such as nanowires~\cite{weber_probing_2010,nadj-perge_spin-orbit_2010,liu_photon_2014}, or employing material systems without piezo electric effect, such as silicon~\cite{mi_strong_2016} or carbon.

\section{Conclusion}
\label{cCon}
We have measured phonon emission rates in a GaAs/AlGaAs \gls{DQD}. The isolation of our system from electronic reservoirs, the low temperature and the weak tunnel coupling compared to values of detuning allowed for a direct readout of the interaction spectral density of electron-phonon coupling. We determined the strength of individual contributions to $\Jint (\varepsilon)$ and found that, apart from the dominant piezoelectric coupling, the deformation-potential coupling becomes important at high energies. The theoretical model captures well the oscillations in the tunneling rate, which arise from the interference of different emission paths. Our results are relevant, for example, for charge qubits, since the observed transition rates explain the dephasing times reported in literature. In addition, we expect our work to provide further insights to photon emission experiments where a considerable background current is visible~\cite{liu_photon_2014,gullans_phonon-assisted_2015,stockklauser_microwave_2015} as well as for spin qubits where the phonons take care of the energy conservation of Zeeman split states~\cite{golovach_phonon-induced_2004,petta_coherent_2005,koppens_driven_2006,golovach_spin_2008,danon_spin-flip_2013}.

\bibliographystyle{aipnum4-1}
\bibliography{bibliography}

\end{document}